\documentclass[twocolumn]{aastex6}

\slugcomment{Manuscript to ApJ}
\shorttitle{The Effects of Accretion Flow Dynamics on the Black Hole Shadow of Sgr A$^{*}$}
\shortauthors{H.-Y. Pu, K.~Akiyama \& K.~Asada}
\begin{document}


\title{The Effects of Accretion Flow Dynamics on the Black Hole Shadow of Sagittarius A$^{*}$}
\author{Hung-Yi Pu\altaffilmark{1}}
\author{Kazunori Akiyama\altaffilmark{2,3}}
\author{Keiichi Asada\altaffilmark{1}}

\altaffiltext{1}{Institute of Astronomy \& Astrophysics, Academia Sinica, 11F of Astronomy-Mathematics Building, AS/NTU No. 1, Taipei 10617, Taiwan}
\altaffiltext{2}{Massachusetts Institute of Technology, Haystack Observatory, Route 40, Westford, MA 01886, USA}
\altaffiltext{3}{National Astronomical Observatory of Japan, 2-21-1 Osawa, Mitaka, Tokyo 181-8588, Japan}

\begin{abstract}
A radiatively inefficient accretion flow (RIAF), which is commonly characterized by its sub-Keplerian nature, is a favored accretion model for the supermassive black hole at Galactic center, Sagittarius A$^{*}$.
To investigate the observable features of a RIAF, we compare the modeled shadow images, visibilities, and spectra of three flow models with dynamics characterized by (i) a Keplerian shell which is rigidly-rotating outside the innermost stable circular orbit (ISCO) and infalling with a constant angular momentum  inside ISCO, (ii) a sub-Keplerian motion, and (iii) a free-falling motion with zero angular momentum at infinity.
At near-mm wavelengths the emission is dominated by the flow within several Schwarzschild radii.
The energy shift due to the flow dynamics becomes important and distinguishable, suggesting that the flow dynamics are an important model parameter for interpreting the mm/submillimeter very long baseline interferometric observations with the forthcoming, fully assembled Event Horizon Telescope (EHT).
As an example, we demonstrate that structural variations of Sagittarius A$^{*}$ on event horizon-scales detected in previous EHT observations can be explained by the non-stationary dynamics of a RIAF.
\end{abstract}

\keywords{accretion, accretion disks --- black hole physics --- Galaxy: center --- submillimeter: general --- techniques: interferometric}

\section{Introduction}
Very long baseline interferometry (VLBI) observations of the nearest supermassive black hole, Sagittarius A$^{*}$ (Sgr A$^{*}$), have provided valuable information concerning the emission region within a few Schwarzschild radii of the black hole \citep[e.g.][]{doe08,fish11,fish16,joh15}. 
The extremely dim luminosity of Sgr A$^{*}$ ($\sim10^{-9} L_{\rm Edd}$, where $L_{\rm Edd}$ is the Eddington luminosity) suggests that the accretion flow around the black hole is in the RIAF regime \citep[e.g.,][]{nar95,man97,oze00,yua03}.
Compared with a cold, geometrically thin, Keplerian rotating disk \citep{nov73,pag74}, the radiative cooling time scale is much longer than the accretion time scale for a RIAF, resulting in a hot, geometrically thick flow with a sub-Keplerian rotation \citep[e.g.][]{ich77,nar94,nar97,qua00}.

With a mass of $\sim4.3\times 10^{6}$~$M_{\odot}$ and at a distance of $\sim8.3$~kpc  \citep{ghe08,gil09a,gil09b,cha15}, Sgr A$^{*}$ is expected to cast a shadow with an angular diameter of $\sim 50$~$\mu$as, as predicted by general relativity \citep{bar73, fal00}.
With ultra-high angular resolution (up to $\sim 20$~$\mu$as) and sufficient sensitivity, the forthcoming mm/sub-mm VLBI observations at $\lambda \lesssim 1.3$~mm ($\nu \gtrsim230$~GHz) with the Event Horizon Telescope (EHT) will obtain the first image of such a shadow region \citep{fish14,lu16}. 
The emission from the accretion flow in the vicinity of the black hole, before reaching a distant observer, will experience considerable energy shifts due to both the motions of the emitting fluid medium and the strong gravity near the black hole \citep{bro04,you12}.
As such, the observed spectrum, image and visibility are important indicators to determine the flow dynamics when the accretion flow within several Schwarzschild radii of the black hole is optically thin enough to be observed (at mm/sub-mm wavelengths in the case of Sgr A$^{*}$), and are therefore important topics to be explored.

In the pioneering work of \citet[][]{fal00}, the authors investigate the observational appearance of the accretion flow and the shadow of Sgr A$^{*}$ at mm/sub-mm wavelengths.
Two kinds of accretion flow dynamics were considered therein: (i) plasma with a free-fall motion, and (ii) plasma modeled as a rigidly-rotating shell in Keplerian motion.
In the latter case, the flow was assumed to be in Keplerian motion outside the ISCO.
Within the ISCO, the flow follows geodesics with energy and angular momentum specified at the ISCO boundary \citep{cun75}. 
Subsequently, by adopting the rigidly-rotating Keplerian shell model, \citet[][]{bro06} further considered separate populations of thermal and non-thermal electrons, based on a radial power-law-dependence of vertically averaged density and temperature profiles found in \citet{yua03}. 
The resulting spectrum is reasonably consistent with the observational data.
Whilst GRMHD simulations provide further detailed, time-dependent modeling of the accretion flow \cite[e.g.][]{nob07,mos09,cha15a,cha15b,dex10,dol12,shc12,shc13,gol16}, (semi-)analytic models enable an efficient and flexible means to survey the vast parameter space, constraining the black hole spin, inclination angle and position angle of Sgr A$^{*}$ \citep[e.g.,][]{bro11a,bro11b,bro16}.

Motivated by the sub-Keplerian nature of RIAFs, in this paper we investigate the resulting spectra, images and visibilities of a black hole surrounded by accretion flows with differing dynamics.
The purpose of this paper is to demonstrate the signature that fluid dynamics alone can produce, through employing a semi-analytic approach. Such an approach will be useful for parameter survey studies for EHT observations and comparison with GRMHD simulation results, as will be applied for our future studies. Of course, these observables are also related to other factors related to the underlying radiative processes, such as magnetic field configuration and the spatial and energy distribution of electrons in the flow.

This paper is organized as follows. In Section 2 we describe how we model the dynamic of a flow, including a flow with sub-Keplerian rotation mimicking a RIAF. In Section 3 we compare and discuss the spectra and images of the flow subject to different dynamics. The visibility amplitudes and EHT observations for Sgr A$^{*}$ in 2009 and 2013 are also compared. Finally, a summary and a discussion of future perspectives are presented in Section 4.

\section{Modeling RIAF Dynamics }
\label{sec:accmodel}

\begin{table*}
\caption{Accretion Flow Model Parameters. }
\label{tab:para}
\begin{centering}
\begin{tabular}{lll} 
\multicolumn{3}{c}{} \\
\multicolumn{3}{c}{$a=0$, $i=68^{\circ}$, $\alpha=\beta=0.5$, $n^{0}_{e,\rm th}=2.5 \times 10^{7}\;{\rm cm}^{-3}$, $T^{0}_{e}=1.5 \times 10^{11}\; {\rm K}$, $n^{0}_{e,\rm nth}=8 \times 10^{4}\;{\rm cm}^{-3}$} \\
\hline
\hline
Model           & $u^{r}$  & $\Omega=u^{\phi}/u^{t}$  \\

\hline

Kep  ($r>r_{0}$)        & $u^{r}_{\rm K}(r)= 0$  & $\Omega_{\rm K}(r)=({r^{3/2}+a})^{-1}$ \\
Kep   ($r\le r_{0}$)       & $ u^{r}_{\rm K}(r)= -[2/(3r_{\rm 0})]^{1/2}[(r_{\rm 0}/r)-1]^{3/2}$  & $\Omega_{\rm K}(r)=(\lambda+aH){[r^{2}+2r(1+H)]}^{-1}$\\
sub-Kep          & $u^{r}_{\rm sub-K}(r,\theta)=u^{r}_{\rm K}+(1-\alpha)(u^{r}_{\rm ff}-u^{r}_{\rm K})$  & $\Omega_{\rm sub-K}(r,\theta)=\Omega_{\rm K}+(1-\beta)(\Omega_{\rm ff}-\Omega_{\rm K})$ \\
free-fall           & $u^{r}_{\rm ff}(r,\theta)=-[2r(r^{2}+a^{2})]^{1/2} \Sigma^{-1}$  & $\Omega_{\rm ff}(r,\theta)=2ar \mathcal{A}^{-1}$ \\

\hline
\end{tabular} \\
\end{centering}
\vspace*{2mm}
Note: $\Sigma\equiv r^{2}+a^{2}\cos^{2}\theta$,
$\Delta\equiv r^{2}-2r+a^{2}$,
$\lambda\equiv (r^{2}_{0}-2a\sqrt{r_{0}}+a^{2})/(\sqrt{r^{3}_{0}}-2\sqrt{r_{0}}+a)$,
$H\equiv (2r-a\lambda)/\Delta$,
$\mathcal{A}\equiv (r^{2}+a^{2}){^2} -a^{2}\Delta\sin^{2}\theta$, where $r_{0}$ is the ISCO radius. In every models, $u^{\theta}=0$ for all $r$.

\end{table*}

\begin{figure*}
\centering{}
\includegraphics[width=0.4\textwidth]{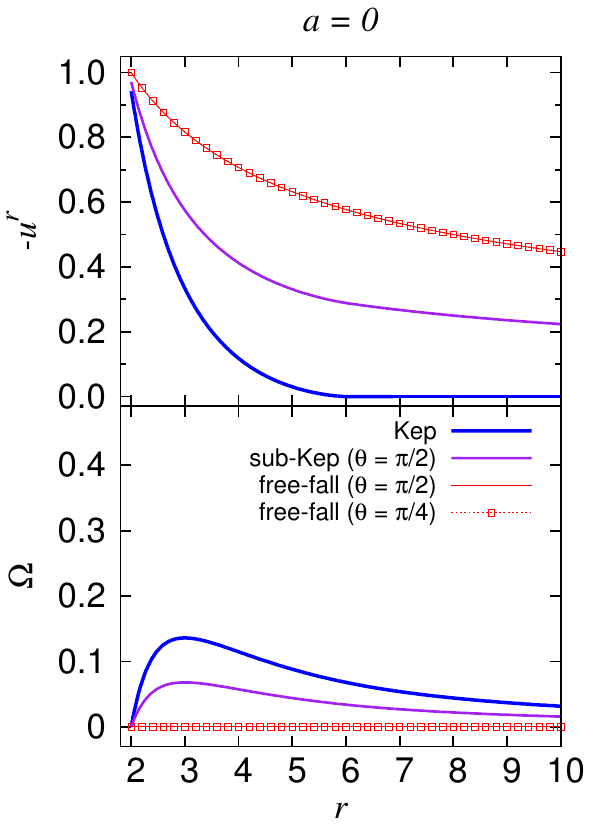}
\includegraphics[width=0.4\textwidth]{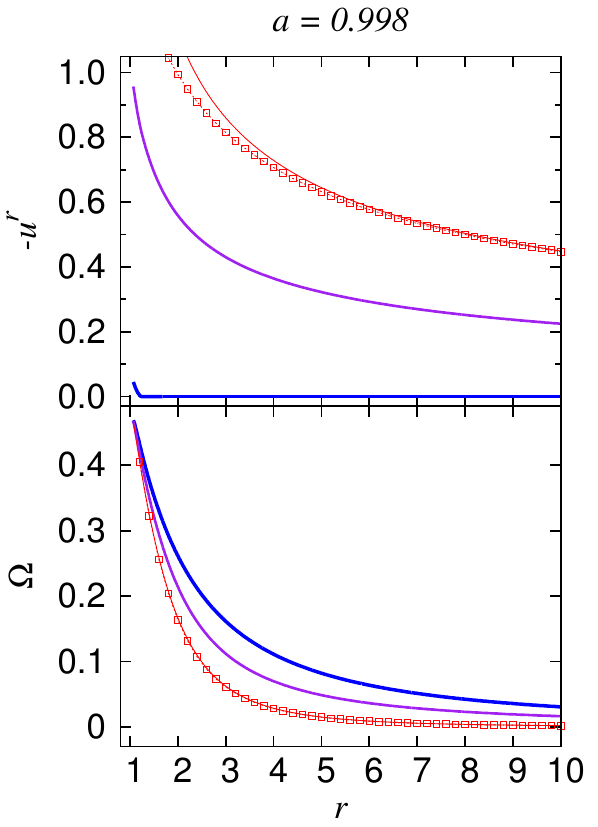}
\caption{Flow dynamics of the three models: Keplerian, sub-Keplerian, and free-fall, for the case $a=0$ (left) and $a=0.998$ (right). The frame-dragging of a spinning black hole causes the $\Omega$ profile to increase as the flow approaches the event horizon. See also Table \ref{tab:para}.}  \label{fig:omega}
\end{figure*}

In order to investigate the sub-Keplerian nature of a RIAF, we consider flow models with dynamics consisting of a combination of stationary free-fall and Keplerian motions \citep[c.f.][]{tak11}. 

Adopting a $[- + + +]$ signature of the background Kerr metric and $c=G=M=1$, the normalization of the four-velocity $u^{\mu}u_{\mu}=-1$ gives 
\begin{equation}\label{eq:model}
(u^{t})^{2}=\frac{1+g_{rr}\left(u^{r}\right)^{2}}{K_{0}}\;,
\end{equation}
where $K_{0}=-\left(g_{tt}+\Omega^{2}g_{\phi\phi}+2\Omega g_{t\phi}\right)$, $\Omega=u^{\phi}/u^{t}$ is the the orbital frequency, and $u^{\theta}$ is assumed to be very small and therefore negligible.  
We can now construct a flow model with a four-velocity $(u^{t},u^{r},u^{\theta}=0,u^{\phi}=\Omega u^{t})$ for a given $\Omega$ through the following steps: (i) determine $u^{r}$, (ii) compute $u^{t}$ from equation (\ref{eq:model}), and then (iii) calculate $u^{\phi}$.
The only physical requirement is that $K_{0}>0$ so that $\left(u^{t}\right)^{2}>0$ is always true. 

It is then straightforward to constrain $u^{r}$ from the Keplerian value $u^{r}_{\rm K} (a, R=r)$ and the free-fall value $u^{r}_{\rm ff} (a, r ,\theta)$, considering the sub-Keplerian case as a combination of the two: 
\begin{equation}
u^{r}_{\rm sub-K} = u^{r}_{\rm K}+(1-\alpha)(u^{r}_{\rm ff} - u^{r}_{\rm K})\;,
\end{equation}
where $a$ is the dimensionless black hole spin parameter, and $0\leqslant\alpha\leqslant1$. 
For the Keplerian disk case, $R$ is the distance to the black hole along the equatorial plane.
A Keplerian rotating shell is thus described by applying $R=r$ (and neglecting the $\theta$--dependence).  

Similarly, $\Omega$ may be constrained by the Keplerian value $\Omega_{\rm K}$ and the free-fall value $\Omega_{\rm ff}$, considering the sub-Keplerian case as the mixture of the two: 
\begin{equation}
\Omega_{\rm sub-K}=\Omega_{\rm K}+(1-\beta)(\Omega_{\rm ff}-\Omega_{\rm K})\;,
\end{equation}
where $0\leqslant\beta\leqslant1$.
The free parameters $\alpha$ and $\beta$ control how much the flow deviates from Keplerian motion ($\alpha=\beta=1$) and from free-fall motion ($\alpha=\beta=0$), specifying the radial and toroidal motion of the flow.
Here we set $\alpha=\beta=0.5$ to represent a sub-Keplerian flow.
This approach is generally applicable for all black hole spins.
The profiles of $u^{r}$ and $\Omega$ for the cases $a=0$ and $a=0.998$ are shown in Fig. \ref{fig:omega}, .
Differences between different flow dynamics become more obvious as the flows approaches the black hole.
This implies that observations at optically thin (i.e. mm/sub-mm) wavelengths are crucial to determine the flow dynamics and discriminate between different models.

In what follows, we adopt the best-fit parameters ($a=0$, inclination angle $i=68^{\circ}$) constrained by the model fitting for 1.3~mm (230~GHz) VLBI observations in \citet{bro11a}, which are also consistent with \citet{bro16}.
We assume hybrid populations of thermal and non-thermal electrons, which obey a relativistic Maxwellian distribution and a power-law distribution in energy, respectively. 
The non-thermal electrons follow a power-law distribution with a spectral index of 1.25 (which corresponds to an energy index of 3.5), and a lower cutoff Lorentz factor of $10^{2}$.
In addition, similarly to \citet{bro11a}, the spatial distributions of the temperature and density of these populations are described by
\begin{eqnarray}
n_{e,\rm th}   &=& n^{0}_{e,\rm th}\;\,r^{-1.1} e^{-z^{2}/2\rho^{2}}\;, \label{eq:p1} \\
T_{e}             &=& T_{e}^{0}\;\,r^{-0.84}\;, \label{eq:p2} \\
n_{e,\rm nth} &=& n^{0}_{e,\rm nth}\;\,r^{-2.02} e^{-z^{2}/2\rho^{2}}\;, \label{eq:p3}
\end{eqnarray}
where $\rho=r \sin\theta$, and $z=r\cos\theta$.
The normalizations of $n^{0}_{e,\rm th}$, $T^{0}_{e}$, and $n^{0}_{e,\rm nth}$, which are equivalent to specifying the accretion rate, are chosen to fit the observed spectrum and are therefore functions of black hole spin and inclination angle.
In such a description, the electron populations decreases rapidly when approaching the funnel region, $\rho\to0$.

The resulting emission is considered as synchrotron emission from the hybrid populations, with  the magnetic field strength determined via, in physical units, 
\begin{equation}
\frac{B^{2}}{8\pi}=0.1\;n_{e,\rm th}\frac{m_{\rm p}c^{2} r_{\rm g}}{6 r}\;,
\end{equation}
where $r_{\rm g}=GM/c^{2}$.
Recent 1.3-mm observations with the EHT have detected linear polarization on scales of several $r_{\rm g}$, indicating the existence of a magnetic field structure which is predominantly stochastic but also partially ordered on event horizon scales \citep[][]{joh15}.
However, while the magnetic field configuration is associated with differential motion within the flow via the coupling between the plasma and the magnetic field, the field configuration resulting from different flow dynamics is unclear.
Here we adopt the angle-averaged emissivity $j_{\rm th}$ in \citet[][]{mah96} for computing thermal synchrotron emission, and the absorption coefficient $\alpha_{\rm th}$ is obtained via Kirchoff's law.
For non-thermal emission, the emissivity $j_{\rm nth}$ and absorption coefficient $\alpha_{\rm nth}$ are taken from \citet[][]{dex12}, by applying a fixed pitch angle of $\pi/3$.
This enables us to ensure all parameters except the flow dynamics remain unchanged.
The resulting images and spectra of the three models, as summarized in Table \ref{tab:para}, are calculated using the general-relativistic radiative transfer code \texttt{Odyssey} \citep{pu16}. 

\section{Results and Discussion}

\begin{figure}
	\centering{}
	\includegraphics[width=0.5\textwidth]{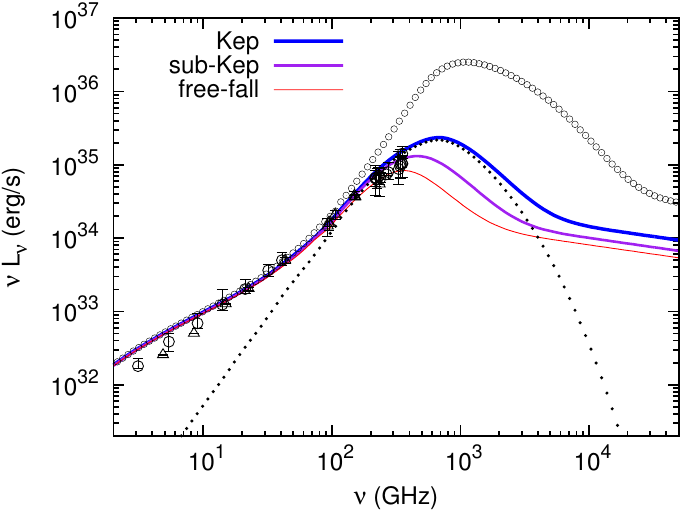}
	\caption{Spectra of three different flow models with different dynamics as shown in the left panel ($a=0$) of Fig. \ref{fig:omega}. Dynamical effects become important when the flow is optically thin at mm/sub-mm wavelengths. Observational data for Sgr A$^{*}$ are taken from \citet{fal98} and \citet{bow15} and are denoted respectively by triangles and circles with corresponding error bars. The dotted line shows the thermal emission for the Keplerian model. The line  with empty circles presents a reference profile when general-relativistic effects are ignored in the radiative transfer calculation. See Section 3 for more details. }
\label{fig:spec}
\end{figure}

\begin{figure*}[t]
\begin{center}
\includegraphics[width=1.\textwidth]{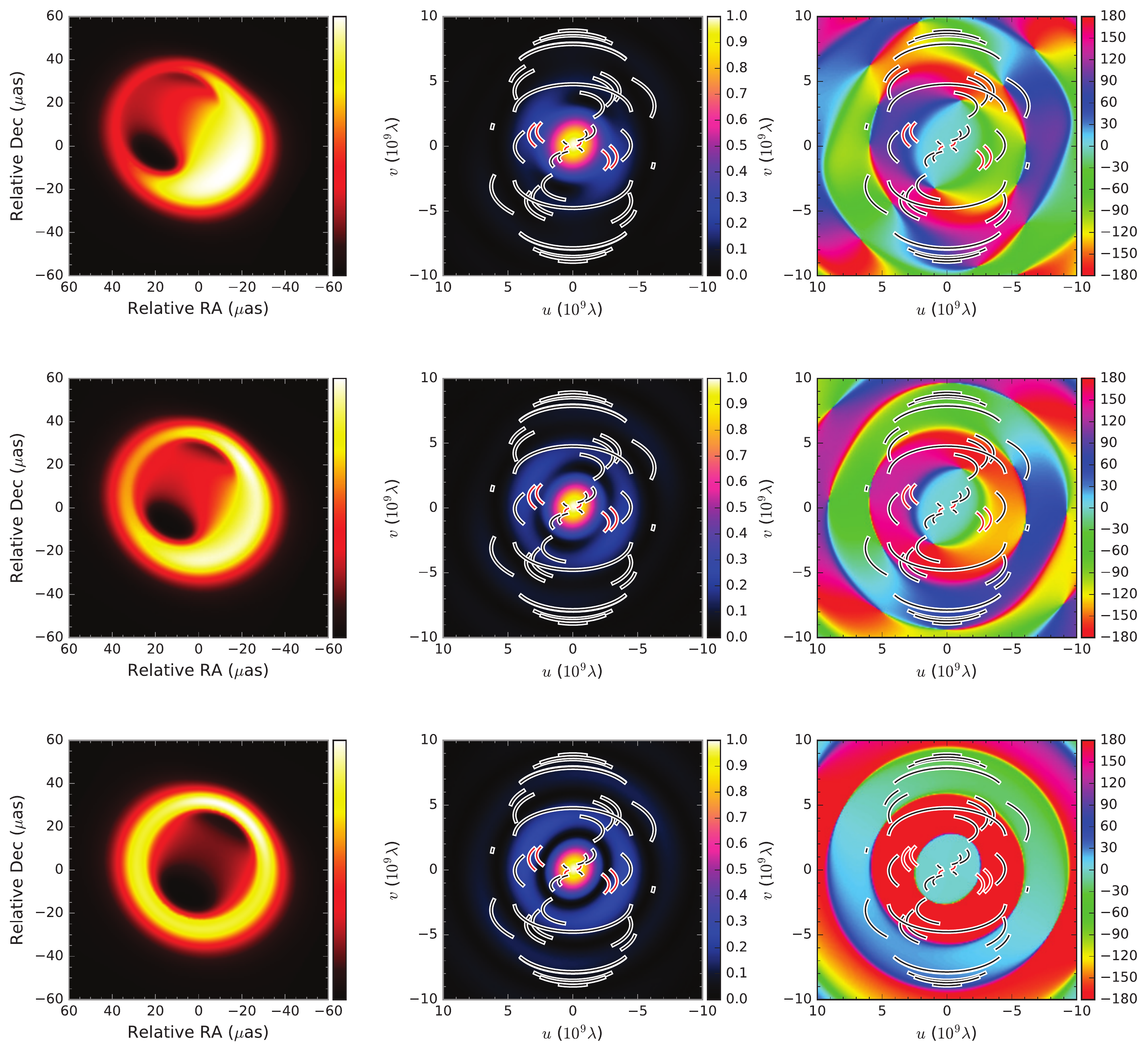}
\caption{The model images at 1.3~mm (230~GHz) (left column), distributions of the visibility amplitudes (central column) and phases (right column) for all three models of the flow dynamics: Keplerian (top row), sub-Keplerian (middle row) and free-fall (bottom row). The images are rotated to a position angle of $150^\circ$, which is close to the best-fit value in \citet{bro16}. The $uv$-coverages of the EHT are shown as red lines for the current array with three US sites, and as black lines for the future full array with an additional five sites \citep[see, e.g.][]{fish14,lu16}. The intensity scale is linear, and normalized for each image (left column),  the visibility amplitudes is normalized by the total flux density (central column), and the phase is in unit of degree (right column).}
\label{fig:image}
\end{center}
\end{figure*}

\begin{figure*}[t]
\centering{}
\includegraphics[width=0.58\textwidth]{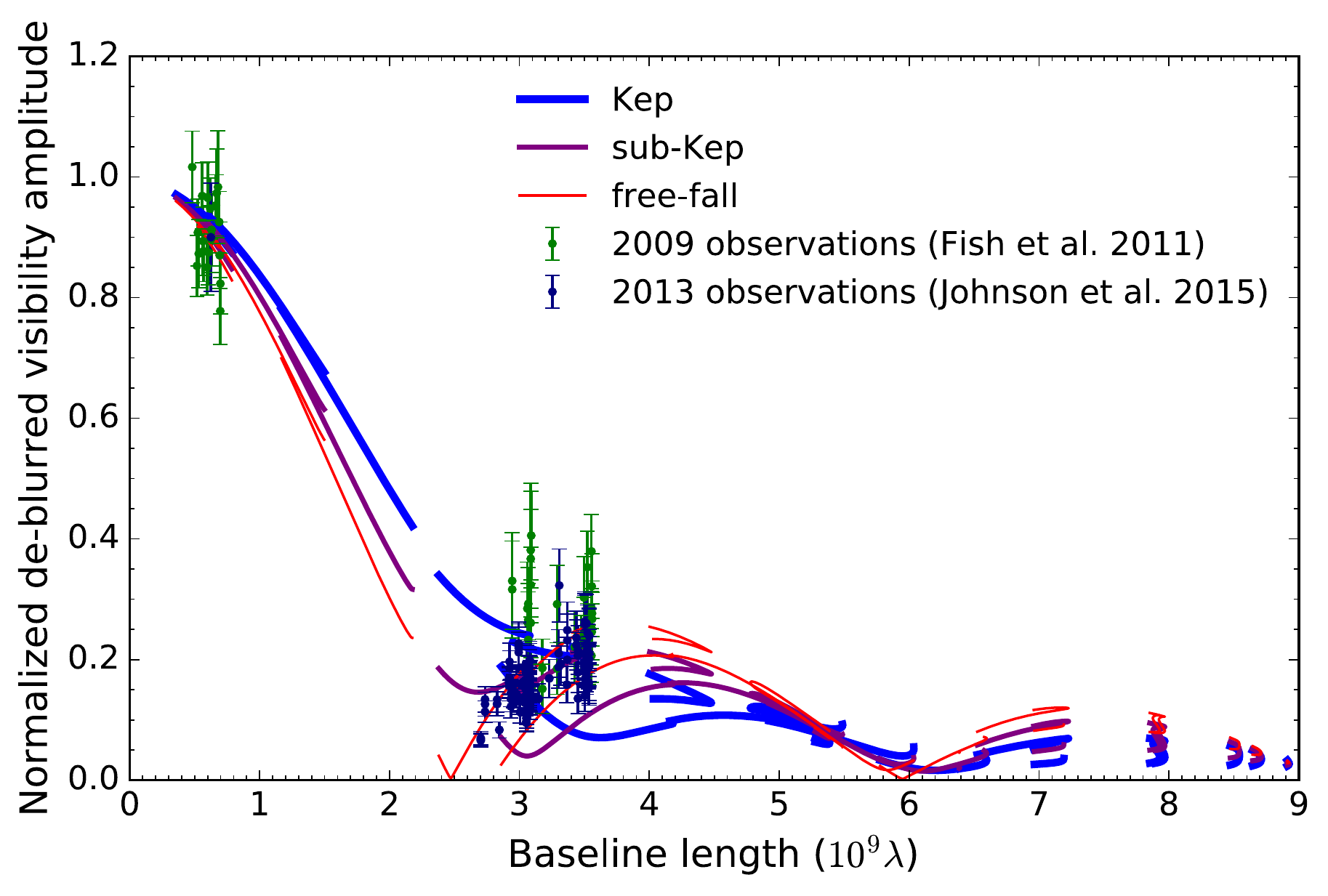}
\caption{
	The visibility amplitude of the three models in Fig. \ref{fig:image} as a function of the baseline length. The solid lines indicate the visibility amplitudes on the $uv$-coverage of the future EHT observations shown in Fig. \ref{fig:image}. For comparison, results of EHT observations in 2009 \citep{fish11} and 2013 \citep{joh15} are shown as purple and navy points, respectively. The angular-broadening due to the scattering effects \citep{jg15} were corrected with the scattering kernel in \citet{bow06}. All amplitude data were normalized so that the average flux density was 0.9 at SMT-CARMA baselines, following \citet{joh15}.
	}
	\label{fig:data}
\end{figure*}

The synchrotron spectra of the three dynamical flow models are presented in Figure \ref{fig:spec} as solid lines, a result of the competition between the emission coefficients and absorption coefficients of the hybrid electron populations in the source function $S=(j_{\rm th}+j_{\rm nth})/(\alpha_{\rm th}+\alpha_{\rm nth})$.
For comparison, the spectrum includes only thermal synchrotron emission ($S=j_{\rm th}/\alpha_{\rm th}$) as shown by dots, for the case of a Keplerian flow.
For all model spectra, thermal synchrotron emission dominates the mm/sub-mm range close to the spectral peak, and non-thermal synchrotron emission is necessary to explain the observed spectrum at lower frequencies. 
 
In order to understand the importance of general-relativistic effects, the empty circles shows the profile for the case where the frequency shift  
\begin{equation}
\frac{\nu}{\nu_{0}}=\frac{p_{\alpha}u^{\alpha}|_{\infty}}{p_{\alpha}u^{\alpha}|_{0}}\; ,
\end{equation}
of the radiation is neglected (i.e.~set to $1$) in the general-relativistic radiative transfer calculation.
In the above equation, $\nu$ is the photon frequency, $p_{\alpha}$ is its covariant four momentum, $u^{\alpha}$ is the four velocity of a fluid particle, and the subscripts ``0'' and ``$\infty$" denotes quantities evaluated, respectively, in the local co-moving frame of the flow and of a distant observer.  
In the lower-frequency region, the surrounding accretion flow remains optically thick and the observed emission is dominated by the photosphere located far from the black hole.
The frequency shift correction is therefore unimportant in this region.
The self-absorbed synchrotron emission in this part of spectrum is dominated by $S=j_{\rm nth}/\alpha_{\rm th}$ \citep[][]{oze00}.
The degeneracy in the dynamics of the three models results in similar spectral profiles.

As the flow becomes optically thin at   several hundred~GHz (i.e. mm/sub-mm wavelengths), the emission from the flow within a few $r_{\rm g}$ of the black hole becomes observable and the black hole shadow is revealed. In general the frequency shift reduces the observed luminosity, shifting the thermal peak to lower frequencies.
The energy shift due to Doppler effects arising from the toroidal motion of the flow results in a larger luminosity in the case of a Keplerian flow compared to other cases with less significant toroidal motion (see also Fig. \ref{fig:omega} for the differences between different flow dynamics).
The profiles of Keplerian and free-fall cases may be taken as the limiting case of a sub-Keplerian RIAF, and therefore represent the boundaries of all possible sub-Keplerian flows (i.e., all possible values of $\alpha$ and $\beta$).
Interestingly, observational data which show variations at different observation epochs are also located within these boundaries.
This indicates that the averaged behavior of non-stationary RIAF flow dynamics at different observation epochs may cause the observed variations. Simultaneous Atacama Large mm/sub-mm Array (ALMA) observations at $345$ GHz, $690$ GHz and $890$ GHz will be extremely useful in constraining these models in the optically thin window ($\gtrsim 100$~GHz).

The corresponding black hole shadow images at $1.3$~mm are dominated by thermal synchrotron emission as shown in the left column of Fig. \ref{fig:image}.
The position angle of the BH spin is set to be $150^\circ$ east of north, close to the best-fit parameter of a Keplerian rotating-flow in \citet{bro16}.
For the case of a Schwarzschild black hole ($a=0$), frame-dragging effects due to the rotation of spacetime are absent, and consequently the luminosity contrast between the left- and right-hand sides of the image (divided by the projection of the rotation axis of the flow) are purely determined by the motion of the flow. 
In Fig. \ref{fig:image} the image of the Keplerian model has the largest luminosity contrast (top left panel), and the image of the free-falling flow is axisymmetric (bottom left panel).
For the case of a Kerr black hole, the resulting shadow image is asymmetric even for a flow in free-fall, with the brighter side corresponding to photons which are Doppler boosted into the observer's line-of-sight by frame-dragging (see e.g. Fig. 1 of \citealt[][]{fal00}; compare also with the $\Omega$ profiles in Fig. \ref{fig:omega}).  Consequently, the resulting crescent structure from the brighter side is the combined effect of both the fluid dynamics and black hole spin in general. 

In Fig. \ref{fig:image}, we also show the distributions of the visibility amplitude and phase, which are Fourier-transformed from the model image and observables of the VLBI, along with $uv$-coverages of future EHT observations with telescopes at eight different sites \citep[see, e.g.,][]{fish14,lu16}.
 In spite of different dynamics, for all cases the visibility amplitudes perpendicular to the spin axis are smoother and more extended compared to the direction parallel to the spin axis. Such characteristics may be used for determining the orientation of the black hole spin, as suggested in \citet{med16}.
In addition, these three models exhibit significantly different visibility distributions not only at baselines longer than $4$~G$\lambda$ traced in future full-array observations, but also at shorter baselines already observed in previous campaigns \citep{doe08,fish11,joh15} with US stations at California (CARMA), Arizona (SMT) and Hawaii (SMA, JCMT and CSO). These baselines are indicated in red in Fig. \ref{fig:image}. 
Therefore the dynamics of a RIAF are an important factor in interpreting both previous EHT observations as well as future observations.

For a more detailed comparison, in Fig. \ref{fig:data} we show the visibility amplitudes of three models on EHT baselines and also the data of EHT observations in 2009 \citep{fish11} and 2013 \citep{joh15}. 
The angular-broadening due to scattering effects \citep{jg15} was corrected with the scattering kernel in \citet{bow06}. 
Interestingly, year-to-year variations in the visibility amplitude can be well-explained simply through the dynamics, without changing other parameters like the black hole spin and inclination angle.
One of the most notable differences between the 2009 and 2013 data is the behavior at long US baselines between the US mainland (SMT, CARMA) and Hawaii, with baseline lengths of $\sim 3$~G$\lambda$\footnote{Note that this could originate from calibration errors, since the gain calibration for the visibility amplitude was improved in the work of \citet{joh15}.}; the amplitude decreased with baseline length in 2009, while it increased in 2013.

The behavior of the 2009 data can be well-explained with the Keplerian model, as already demonstrated in previous studies \citep{bro11a,bro11b,bro16}. In the case of the Keplerian model, since the black hole shadow is partially covered and smoothed by emission from the approaching side of the toroidal flow, the visibility amplitude has a Gaussian-like behavior in the NEE-SWW direction, where long US baselines are distributed (see top middle panel of Fig. \ref{fig:image}).
On the other hand, the rising amplitudes in the 2013 data suggest that the black hole shadow should be more clear than in previous years, which can be well reproduced with sub-Keplerian or free-fall models.
This indicates that year-to-year structural variations in Sgr A$^{*}$ on event-horizon scales potentially reflect the non-stationary dynamics of a RIAF, whilst simultaneously suggesting that consideration of the flow dynamics is essential in understanding and interpreting data from mm/sub-mm VLBI observations.

\section{Summary and Future Perspective}
We considered three different accretion flow models: (i) a Keplerian shell rotating outside the ISCO and infalling with constant angular momentum inside the ISCO (e.g. \citealt[][]{fal00,bro06}), (ii) a sub-Keplerian flow (to mimic the flow motion of a RIAF), and (iii) a flow in free-fall with zero angular momentum at infinity (e.g. \citealt{fal00}).
It was demonstrated that the sub-Keplerian nature of a RIAF is important for modeling black hole images at mm/sub-mm wavelengths, and hence crucial for interpreting the observed VLBI visibility and variability of Sgr A$^{*}$. 

 Our semi-analytic model of RIAF dynamics are constructed through a combined description of both Keplerian and free-fall fluid motion, and are therefore everywhere sub-Keplerian. 
As first noted by \citet[][]{abr81}, if a hydrodynamical flow has a radial profile which is everywhere sub-Keplerian, plasmas in the vicinity of the black hole are then able to flow into higher latitudes near the horizon, $r_{\rm h}$, resulting in a quasi-spherical geometry.  That is, the height $H$-to-cylindrical radius $R$ ratio follows a profile $(H/R)|_{r\gg r_{\rm h}}\sim1$ and $(H/R)|_{r_{\rm h}}\sim1$, which is consistent with the assumption of the spatial populations described in equations (\ref{eq:p1}) and (\ref{eq:p3}).  It should be noted that we assume our RIAF model characterises a stationary flow state for given values of $\alpha$ and $\beta$ (which respectively control $u^{r}$ and $\Omega=u^{\phi}/u^{t}$ of the flow), and that we assume these values may vary from observation to observation due to the sufficiently long dynamical time scale between observations. In addition, magnetic effects are ignored.
The simple model is useful for future exploration of the parameter space,  providing a quick, flexible overview of the most likely flow dynamics ranges.
We plan to perform further detailed analysis using the closure phase data recently published in \citet{fish16}, which will prove useful to constrain the dynamics of the accretion flow as well as the black hole spin, its inclination angle, and position angle.

 While GRMHD simulations with an initial equilibrium torus ($(H/R)|_{r\gg r_{\rm h}}\sim1$) have successfully simulated sub-Keplerian accretion flows when a stationary state is reached \citep[e.g.,][]{mck12,nar12}, further examination of how $(H/R)|_{r_{\rm h}}$ is affected by the presence of a magnetic field will be important to understand the spatial distribution of plasma in the very vicinity of the event horizon. 
 For example, it has been discussed in \citet[][]{mck12} that the accumulated polar magnetic flux can compress the flow height when approaching the horizon, resulting in $(H/R)|_{r_{\rm h}}\ll1$ (see top panel of Fig. 11 in \citealt[][]{mck12} and Fig. 1 in \citealt[][]{ori16}). Such an understanding is directly related to the shadow image when the flow becomes optically thin.
In addition, compared to that of a simulation in which an initial equilibrium torus is adopted, it is expected that the resulting fluid dynamics of a simulation in which mass is continuously injected onto the grid would be more closer to a free-fall scenario. Such boundary conditions have been applied in many hydrodynamical or MHD simulations with a pseudo-Newtonian potential \cite[][]{mck02,igu00,igu02,igu03}. Future comparison between different GRMHD simulation results and our model parameters $\alpha$ and $\beta$ will help to improve the modelling of RIAF flows.

\acknowledgments
\label{ack}
We thank Ziri Younsi for helpful discussions and proof reading. We also thank Vincent Fish and Roman Gold for comments. H.-Y. Pu and K. Asada  are supported by the Ministry of Science and Technology (MOST) of Taiwan under the grant MOST 103-2112-M-001-038-MY2. K. Akiyama is financially supported by the program of Postdoctoral Fellowships for Research Abroad at the Japan Society for the Promotion of Science and grants from the National Science Foundation (NSF). This research has made use of NASA's Astrophysics Data System.


\end{document}